\documentclass[preprint,showpacs,preprintnumbers,amsmath,amssymb]{revtex4}

\usepackage{graphicx}
\usepackage{dcolumn}
\usepackage{bm}

\begin{document}

\preprint{}

\title{Thermal Pions at Finite Isospin Chemical Potential}

\author{Marcelo Loewe}
\author{Cristi\'an Villavicencio}%
\affiliation{%
 Facultad de F\'{\i}sica,
 Pontificia Universidad Cat\'olica de Chile,
 Casilla 306, Santiago 22, Chile}%

\begin{abstract}
The density corrections, in terms of the isospin chemical
potential $\mu_I$, to the mass of the pions are studied in
the framework of the $SU(2)$ low energy effective chiral
lagrangian. The pion decay constant $f_{\pi }(T, \mu _{I})$ is also analized.
As a function of temperature for $\mu_I =0$, the mass
remains quite stable,  starting to grow for very high
 values of $T$,  confirming previous results. However,
  there are interesting corrections to the mass when
both effects (temperature and chemical potential) are
simultaneously present. At zero temperature the $\pi ^{\pm}$
 should condensate when $\mu _{I} = \mp m_{\pi }$. This is not
 longer valid anymore at finite $T$. The mass of the $\pi_0$
 acquires also a non trivial dependence on $\mu_I$ due to the  finite temperature.
\end{abstract}


\maketitle

Pions play a special role in the dynamics of hot hadronic matter
since they are the lightest hadrons. Therefore, it is quite
important to understand not only the temperature dependence of the
pion's Green functions but also their behavior as function of
density, through the chemical potential. The dependence of the
pion mass and decay constant on temperature $m_{\pi}(T)$,
$f_\pi(T)$ has been studied in a variety of frameworks, such as
thermal QCD-Sum Rules \cite{DFL}, Chiral Perturbation Theory (low
temperature expansion) \cite{GL}, the Linear Sigma Model
\cite{LCL}, the Mean Field Approximation \cite{Bar1}, the Virial
Expansion \cite{Schenk}, etc. In fact, the pion propagation at
finite temperature has been calculated at two loops in the frame
of chiral perturbation theory \cite{Schenk2,T}. There seems to be
a reasonable agreement that $m_{\pi}(T)$ is essentially
independent of $T$, except possibly near the critical temperature
$T_{c}$ where $m_{\pi}(T)$ increases with $T$ and that $f_\pi(T)$
vanishes for the critical temperature.

The introduction of in-medium processes via isospin chemical
potential has been studied at zero temperature
\cite{son,toublan,splittorff} in both phases ($|\mu_I|\lessgtr
m_\pi$) at tree level.
 The problem with both, temperature and density, has been worked
 out for barionic
 chemical potential with Chiral Perturbation Theory \cite{AE-N}. It is also possible to
 find certain region of the stable pion gas in which the pion number is locally
 conserved \cite{ayala}.\\

 Usually, there are two
procedures to extract the information of $m_\pi$ and $f_\pi$ in
the frame of chiral perturbation theory. The first one  is to
compute the Axial-Axial correlator which provides us with the
decay constant and the mass corrections. \cite{GL,GL2,Schenk2}
\begin{equation}
\int d^4x e^{ip\cdot x}\langle 0|A_\mu^a(x)A_\nu^b(0)|0\rangle
=\delta^{ab}\frac{p_\mu p_\nu f_\pi^2}{p^2-m_\pi^2}
\end{equation}
In the second method, radiative corrections to the propagators are
considered together with the realization of PCAC, $\langle
0|A_\mu^a|\pi^b(p)\rangle=ip_\mu \delta^{ab}f_\pi$ making then use of
appropriate counterterms.
 The
use of counterterms is not necessary in the Axial-Axial
correlator method. We have checked that both methods leave the same answers.
 Let us proceed in the frame of the $SU(2)$ chiral
perturbation theory. The most general chiral invariant expression
for a QCD-extended lagrangian, \cite{GL2,pich} under the presence
of external hermitian-matrix auxiliary fields, has the form
\begin{eqnarray}
{\cal L}_{\tiny QCD}(s,p,v_\mu,a_\mu)   &=&  {\cal L}_{\tiny QCD}^0\nonumber\\
                                        &+& \bar q \gamma^\mu (v_\mu +\gamma_5 a_\mu)q\nonumber\\
                                        &-& \bar q(s-i\gamma_5
                                        p)q,
\end{eqnarray}
where  $v_\mu$, $a_\mu$, $s$ and $p$ are vector, axial, scalar
and pseudoscalar fields. The vector current  is given by
\begin{equation}
J_\mu^a=\frac{1}{2}\bar q \gamma_\mu \tau^aq.
\end{equation}

When  $v,a,p=0$ and $s=M$, being $M=diag(m_u,m_d)$ the mass
matrix, we obtain the usual $QCD$ lagrangian. This procedure is
formal, in the sence that we reproduce the usual QCD lagrangian with current masses.
However, we would like to notice that a scalar field in chiral lagrangian
models the spontaneus break of chiral symmetry through a non vanishing vacuum
expectation value. In this sense if we take for $s=M$, these masses should be actually
constituent quark masses, while in the QCD lagrangian we have current masses.
Nevertheless this is a formal step which tries only to motivate what follows in the
context of effective pion lagrangian.

 The effective action with
finite isospin chemical potential is given by
\begin{eqnarray}
{\cal L}_{\tiny QCD}^I    &=& {\cal L}_{\tiny QCD}(M,0,0,0)+\mu^a u^\mu  J^a_\mu\nonumber\\
                    &=& {\cal L}_{\tiny QCD}(M,0,\mu u_\mu,0) \label{eq:accion}
\end{eqnarray}
where $\mu^a=(0,0,\mu_I)$ is the third isospin component,
$\mu=\mu^a\tau^a/2$ and $u_\mu$ is the 4-velocity between the
observer and the thermal heat bath. This is required in order to describe
in a covariant way this system, where the Lorentz invariance is
broken since the thermal heath bath represents a privileged frame
of reference.

Proceeding in the same way, now  in the low-energy description where
only pion degrees of freedom are relevant, let us consider the
most general chiral invariant lagrangian
ordered in a series of powers of the external momentum.
We will start with the ${\cal O}(p^2)$ chiral lagrangian
\begin{eqnarray}
    {\cal L}_2  &=& \frac{f^2}{4}Tr\left[(D_\mu U)^\dag D^\mu U+U^\dag \chi
+ \chi^\dag U\right]
\end{eqnarray}
with
\begin{eqnarray}
 D_\mu U &=& \partial_\mu U-i[v_\mu, U]-i\{ a_\mu, U\}\nonumber\\
 \chi &=& 2B(s+ip)\nonumber\\
 U &=& \bar U^\frac{1}{2}(e^{i\pi^a\tau^a/f})\bar U^\frac{1}{2}
 \label{eq:fields}
\end{eqnarray}
$\bar U$ is the vacuum expectation value of the field $U$ and
$B$ in the previous equation is an arbitrary constant which
will be fixed when the mass is identified setting $(m_{u} +
m_{d})B =m^{2}$.
 The most general ${\cal O}(p^4)$ chiral lagrangian has the form

\begin{eqnarray}
{\cal L}_4 &=&  \alpha_1\left( Tr\left[(D_\mu U)^\dag D^\mu U\right]\right)^2\nonumber\\
           &+&  \alpha_2Tr\left[(D_\mu U)^\dag D_\nu U\right]Tr\left[(D^\mu U)^\dag D^\nu U\right]\nonumber\\
           &+&  \alpha_3\left(Tr\left[\chi U^\dag
                                        +U\chi^\dag\right]\right)^2\nonumber\\
           &+&  \alpha_4Tr\left[(D_\mu U)^\dag D^\mu U\right]Tr\left[\chi U^\dag + U\chi^\dag    \right]\nonumber\\
           &+&  \alpha_5\left[L_{\mu\nu}UR^{\mu\nu}U^\dag\right]\nonumber\\
           &+&  i\alpha_6Tr\left[L_{\mu\nu}D^\mu U(D^\nu U)^\dag +R_{\mu\nu}(D^\mu U)^\dag D^\nu U\right]\nonumber\\
           &+&  \alpha_7\left(Tr\left[\chi U^\dag-U\chi^\dag\right]\right)^2\nonumber\\
           &+&  \alpha_8Tr\left[\chi U^\dag \chi U^\dag +U\chi^\dag U\chi^\dag\right]\nonumber\\
           &+&  \alpha_9Tr\left[L_{\mu\nu}L^{\mu\nu}+R_{\mu\nu}R^{\mu\nu}\right]\nonumber\\
           &+&  \alpha_{10}Tr[\chi^\dag\chi]
\end{eqnarray}
with
\begin{eqnarray}
    L_{\mu\nu}  = \partial_\mu l_\nu -\partial_\nu l_\mu +i[l_\mu,l_\nu],
                    &&\quad l_\mu       = v_\mu -a_\mu\nonumber\\
    R_{\mu\nu}  = \partial_\mu r_\nu -\partial_\nu r_\mu +i[r_\mu,r_\nu],
                    &&\quad r_\mu       = v_\mu +a_\mu
\end{eqnarray}

The different coupling constants $\alpha _{i}$in the previous expression are related to
the couplings introduced by \cite{GL1}. Here we use the prescription of \cite{Scherer}.

The effective action with finite chemical potential in terms of
pion degrees of freedom has the same form as eq.\ref{eq:accion},
where the different external fields are defined in
eq.\ref{eq:fields}. In this paper we will consider one loop
corrections, up to the fourth order in the fields, to the
lagrangian ${\cal L}_2$ and the free part, i.e the tree level part
of ${\cal L}_4$ with renormalized fields. This procedure is
standard, \cite{GL2,holstein}. We will concentrate on the phase where $\mu _{I} < m_{\pi }$, where
the vacuum expectation value $\bar{U} =1$.
 The interacting part ${\cal L}_{4}$ involves higher powers in the momentum  of the pion fields.
  The constants $\alpha _{i}$ present in ${\cal L}_{4}$ are known  from decay and scattering measurements. Therefore, we have the following lagrangians ${\cal L}_{i,j}$
\begin{eqnarray}
    {\cal L}_{2,2} &=&
        \frac{1}{2}\left[\left(\partial\pi_0\right)^2-m^2\pi_0^2\right]
        +\left|\partial_I\pi\right|^2-m^2\left|\pi\right|^2\nonumber\\
{\cal L}_{2,4} &=& \frac{1}{4!}\frac{m^2}{f^2}\pi_0^4 +
\frac{1}{6f^2}\left[
-4\left|\partial_I\pi\right|^2\left|\pi\right|^2
\right.   \nonumber\\
               & &  \qquad\qquad\qquad\left.  +\left(\partial\left|\pi\right|^2\right)^2 + m^2\left(\left| \pi\right|^2\right)^2\right]\nonumber\\
                & & +\frac{1}{6f^2}\left[ -2\left|\partial_I\pi\right|^2\pi_0^2
                                            -2\left(\partial\pi_0\right)^2\left|\pi\right|^2
                                            \right.\nonumber\\
                & &  \qquad \quad    \left.   +\partial\pi_0^2\cdot\partial\left|\pi\right|^2
                +m^2\pi_0^2\left|\pi\right|^2\right]\nonumber\\
{\cal L}_{4,2}   &=&
2\frac{m^2}{f^2}\left[l_4\left|\partial_I\pi\right|^2
                                         -m^2(l_3+l_4)\left|\pi\right|^2\right.\nonumber\\
                    & &       +  \left.\frac{1}{2}l_4\left(\partial\pi_0\right)^2
                        -\frac{1}{2}m^2(l_3+l_4-\epsilon_{ud}^2l_7)\pi_0^2\right]
\label{eq:a,b}
\end{eqnarray}
with $l_i$ the original parameters of Gasser \& Leutwyler $SU(2)$ lagrangian
\begin{eqnarray}
l_4   &=& 8\alpha_4\nonumber\\
l_3   &=& 16\alpha_3+8\alpha_8-8\alpha_4\nonumber\\
l_7   &=& -16\alpha_7-8\alpha_8
\end{eqnarray}
where the subindexes ($i,j$) in the lagrangian denote the order in
powers of momentum and  fields, respectively, and
 $$\partial_{I\pm}\equiv
\partial\pm i\mu_Iu.$$
This definition of the covariant derivative is
natural, since we know
\cite{weldon,actor}  that the chemical potential is introduced as
the zero component of an external ``gauge" field. In the previous
expression, $$|\pi|^2 \equiv \pi^+\pi^- , \qquad
|\partial_I\pi|^2=(\partial_I \pi)_+(\partial_I \pi)_-.$$ We will neglect $\epsilon
_{ud}^2  = (m_u-m_d)^{2}/(m_u + m_d)^{2}$
because it only shifts in a small quantity the neutral pion mass
and we are interested in the thermal and density evolution of the
masses.

For renormalizing with counterterms we introduce the following
decomposition
\begin{eqnarray}
    {\cal L}_{eff}  &=& {\cal L}_{2,2} + {\cal L}_{2,4}^{r}+{\cal L}^{r}_{4,2}\nonumber\\
     {\cal L}_{2,2}   &=& {\cal L}^{r}_{2,2}+\delta{\cal L,}
\end{eqnarray}
where the $r$ index denote the lagrangian with renormalized
fields.

Setting $\pi_0=\sqrt{Z_0}\pi_0^r$ and
$\pi_\pm=\sqrt{Z_\pm}\pi_\pm^r$ in ${\cal L}_{2,2}$, we have
\begin{eqnarray}
\delta{\cal L} &=&
\frac{1}{2}\delta_{Z_o}\left[(\partial\pi_0^r)^2
                    -m^2(\pi_0^r)^2\right]\nonumber\\
                 & &   +\delta_{Z_\pm}\left[\left|\partial_I\pi^r\right|^2
                    -m^2\left|\pi^r\right|^2\right]
\end{eqnarray}
with $\delta_{Z_i}=Z_i-1$.

First, let us consider the temperature and density corrections to the
pion propagator. Since our
calculation will be at the one loop level, we do not need the full
formalism of thermo field dynamics, including thermal ghosts and
matrix propagators. The propagator
\begin{equation}
D_\pm (x)=D(x;\pm\mu_I)+D_\beta ( x;\pm\mu_I)
\end{equation}
 for charged
pions at the tree level will be given by an extension, for a
non-vanishing chemical potential, of the well known Dolan-Jackiw
propagators for scalar fields \cite{weldon}. Note that since there
is no chemical potential associated to the neutral pion, the
thermal propagator $D_{0}$ will be the usual one
\begin{eqnarray}
D_0(x)=D(x;0)+D_\beta (x;0)
\end{eqnarray}
 where, in momentum space
\begin{eqnarray}
\underline{\mathrm D}(k;\pm\mu_I)  &=& \frac{i}{k_\pm^2-m^2+i\epsilon}\nonumber\\
\underline{\mathrm D}_\beta (k;\pm\mu_I)&=& 2\pi n_B(|k\cdot u|)\delta(k_\pm^2-m^2)
\end{eqnarray}
with
$$k_\pm\equiv k\mp\mu_Iu, \qquad n_B(x)=\frac{1}{e^{\beta x}-1}$$
are the shifted momentum and the Bose-Einstein factor.

\bigskip
\noindent

 We will use the $\overline{MS}$-scheme, and we renormalize as usual at $T=0$, since the thermal corrections
are finite. The self energy for charged  and neutral pions including the
counterterms has the form
\begin{eqnarray}
\Sigma_\pm(p)  &=&[A_\pm-\delta_{Z_\pm}]p_\pm^2-[A_\pm'- \delta_{Z_\pm}]m^2
+A_\pm'' u\cdot p_\pm \nonumber\\
\Sigma_0(p)&=& [A_0-\delta_{Z_0}]p^2-[A_0'-\delta_{Z_0}]m^2
\end{eqnarray}
with
\begin{eqnarray}
A_\pm  &=&  \frac{1}{3f^2}\left[ D_\pm (0)+ D_0(0)\right]-2\frac{m^2}{f^2}l_4\nonumber\\
A_\pm '  &=&  \frac{1}{6f^2}\left[2 D_\pm (0)- D_0(0)\right]-2\frac{m^2}{f^2}(l_3+l_4)\nonumber\\
A_\pm ''    &=& \frac{2}{f^2}u\cdot\partial D_\pm(0)\nonumber\\
A_0 &=& \frac{2}{3f^2}D_\pm (0)-\frac{m^2}{f^2}l_4\nonumber\\
A_0'    &=& \frac{1}{6f^2}\left[3D_0 (0)-2D_\pm (0)\right] -\frac{m^2}{f^2}(l_3+l_4)
\end{eqnarray}

 Our
prescription to fix the counterterm $\delta_{Z_\pm}$ is to impose
that $\Sigma$ does not depend on $p^2$, so, $\delta_{Z_i}=A_i$. In
this way,  the renormalized propagators will take the form
\begin{eqnarray}
i\underline{\mathrm D}_\pm^r(p)^{-1}&=&p_\pm^2-A_\pm ''u\cdot p_\pm -m^2[1-A_\pm '+A_\pm]\nonumber\\
i\underline{\mathrm D}_0^r(p)^{-1}&=& p^2-m^2[1-A_0'+A_0]
\end{eqnarray}

 where $\alpha_i$ terms absorbs the divergences
\begin{eqnarray}
\l_i
    &=& \frac{\gamma_i}{32\pi^2}\left[\bar\l_i+\ln\frac{4\pi m^2}{\Lambda^2}-\frac{2}{d-4} -\gamma +1\right]
\end{eqnarray}
 in which the $\gamma_i$
terms are tabulated \cite{GL2,holstein}, being $\Lambda$ a scale factor.

We identify $m_{\pi^+}$ and $m_{\pi^-}$
from the solution of $\underline{\mathrm D}_\pm^r(p)^{-1}|_{{\bf p}=0}=0$
in the frame where the
heath bath is at rest ($u=(1,{\bf 0})$).
We get the well known result
for $T=\mu_I=0$
\begin{equation}
m_\pi =m \left(
    1-\alpha_\pi \bar l_3/4
        \right)
\end{equation}
is identified with the physical mass.
 $\alpha_\pi=(m_\pi/4\pi f_\pi)^2$ is the perturbative term that
fixes  the scale of energies in the theory (for energies below
$4\pi f_\pi $) so we neglect the ${\cal O}(g^2)$ factors. This
allows us to set $m \approx m_\pi$ in all radiative corrections
(and also $f \approx f_\pi$). The procedure is the same for
$m_{\pi^0}$

It is important to remark that radiative corrections will leave a
dependence on the chemical potential for the pion mass only for
finite values of temperature. In a strict sense, this procedure
does not allow us to say nothing new for an eventual chemical
potential dependence of the masses at $T=0$ (cold matter) which is
already included in ${\cal L}_2$. In this case, $T=0$, we have to
follow the usual procedure, \cite{son,toublan}, of computing the
minimum of the effective potential in ${\cal L}_2$ when the
chemical potential is taken into account, without considering
radiative corrections. This enables to identify a phase structure
where a non trivial vacuum appears for higher values of $\mu
_{I}$, $|\mu _{I}| > m_{\pi }$characterized by the appearance of a condensate $\langle
\pi ^{-}\rangle$. (The opposite occurs for negative values of the
chemical potential, where  the vacuum state is a condensate
$\langle \pi ^{+} \rangle $). At $T=0$
when $\mu _{I} = m_{\pi }$, the mass of $\pi ^{-} $ vanishes.

For finite $T$ and $\mu_I$, we find the following expression for
the masses
\begin{eqnarray}
m_{\pi^\pm}(T,\mu_I)\!\! &=&  \!\!  m_\pi\left[1+\alpha_\pi
I_{0}\pm\left(\mu_I/m_\pi-4\alpha_\pi J\right)\right]\nonumber\\
m_{\pi^0}(T,\mu_I)\!\! &=&\!\!
m_\pi\left[1+\alpha_\pi\left(2I-I_{0}\right)\right]
\end{eqnarray}
with
\begin{eqnarray}
I   &=& \int_1^\infty dx\sqrt{x^2-1}
        [n_B(m_\pi x-\mu_I)+n_B(m_\pi x+\mu_I)]\nonumber\\
J   &=& \int_1^\infty dxx\sqrt{x^2-1}
        [n_B(m_\pi x-\mu_I)-n_B(m_\pi x+\mu_I)]\nonumber\\
I_0 &=& I(\mu_I=0)\label{ints}
\end{eqnarray}

Note that our convention for the chemical potential sign is contrary to
the one adopted in the paper by Kogut and Toublan \cite {toublan},
 who extended previous results by Son and Stephanov \cite{son}.

If the chemical potential of the charged pions vanishes, i.e for
symmetric matter,  at finite T we get the well known result for
$m_{\pi }(T)$ due to chiral perturbation theory \cite{GL}, see
also \cite{LCL}.
 However, due to radiative corrections to the neutral pion propagator,
  its mass will acquire a non trivial chemical potential dependence for finite values of temperature.
   In the approach where the minimum of the effective potential is calculated (for finite $\mu _{I}$ and $T=0$),
    the mass of the neutral pion remains constant.

We show in Fig.1 a tridimensional picture for the behavior of the
mass of the neutral pion.
 Note that when $\mu_I=0$, $m_{\pi_0}(T)=m_{\pi^\pm}(T)$.
\begin{figure}
\includegraphics{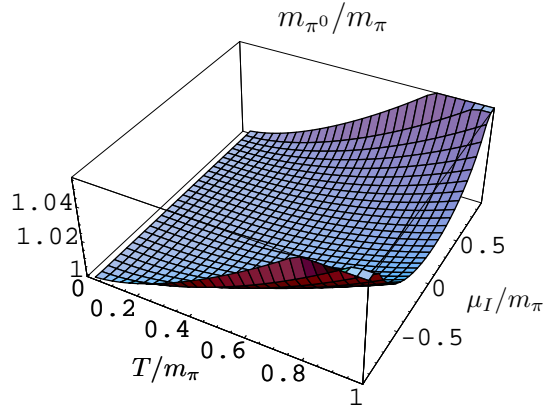}
\caption{$m_{\pi^0}$ as function of $T$ and $\mu_I$ in units of
$m_\pi$ } \label{fig1}
\end{figure}

From Fig.2 we see that at zero temperature, we agree with the
usual prediction, $m_{\pi }^{+} = m_{\pi } + \mu _{I}$. In fact,
at zero temperature the $\pi ^{+}$ should condensate when $\mu
_{I} = -m_{\pi }$ (the inverse situation occurs for $\pi ^{-}$).
 Now, this situation changes if temperature starts to grow. The
 condensation point disappear at $\mu_I=-m_\pi$; in $\mu_I=m_\pi$ the mass start to decrease.
 For small $T$ (for example inside an neutron star),
  this effect is neglegible.
\begin{figure}
\includegraphics{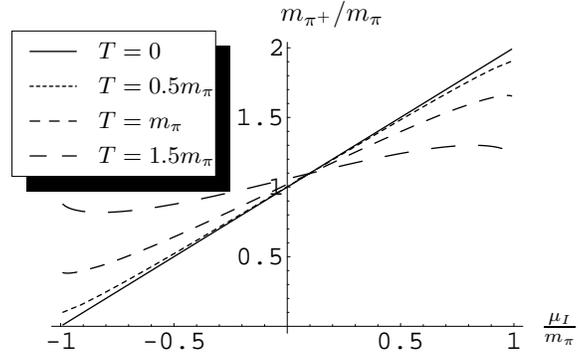}
\caption{ $m_{\pi^+}$ as function of  $\mu_I$ for a fixed $T$ }
 \label{fig2}
\end{figure}

In connection with the behavior of $f_{\pi }(T,\mu _{I} )$ when $\mu _{I} < m_{\pi }$, we have make used of
PCAC, which provides us with a relation between the renormalized propagator and the pion decay constant.

The Axial current is obtained as the functional derivative of the action with respect to $a_\mu^a$, with $a_\mu=a_\mu^a\tau^a/2$
\begin{equation}
A_\mu^a = \frac{\delta{\cal S}}{\delta a_\mu^a}(M,0,\mu u_\mu,0)
\end{equation}

The axial current is
\begin{eqnarray}
A_{(1,1)\mu}^\pm  &=& -f(\partial_\mu^I\pi)^\pm\nonumber\\
A_{(1,3)\mu}^\pm    &=&
\frac{2}{3f}\left\{\pi^0\left[\pi^0(\partial_\mu^I\pi)^\pm
-\pi^\pm\partial_\mu
\pi^0\right]\right.\nonumber\\
    && \left. +\pi^\pm\left[\pi^\mp(\partial_\mu^I\pi)^\pm-\pi^\pm(\partial_\mu^I\pi)^\mp\right]\right\}\nonumber\\
A_{(3,1)\mu}^\pm    &=& -\frac{m^2}{f}2l_4
(\partial_\mu^I\pi)^\pm\nonumber\\
A_{(1,1)\mu}^0   &=&    -f\partial_\mu\pi^0\nonumber\\
A_{(1,3)\mu}^0    &=&   \frac{2}{3f}\left\{
2|\pi|^2\partial_\mu\pi^0-\pi^0\partial_\mu|\pi |^2\right\}\nonumber\\
A_{(3,1)\mu}^0  &=& -\frac{m^2}{f}2l_4 \partial_\mu\pi^0
\end{eqnarray}
 Now, the effective axial current
at ${\cal O}(p^3)$ will be
\begin{eqnarray}
A_\mu^i    &\equiv&
A_{(1,1)\mu}^{i}+A_{(1,3)\mu}^{ir}+A_{(3,1)\mu}^{ir}\nonumber\\
    &=& \sqrt{Z_i}A_{(1,1)\mu}^{ir}+A_{(1,3)\mu}^{ir}+A_{(3,1)\mu}^{ir}
\end{eqnarray}
with $i=\{\pm,0\}$. We will take
$$\sqrt{Z_i}=\sqrt{1+\delta_{Z_i}}\simeq
1+\frac{1}{2}\delta_{Z_i}+{\cal O}(\delta_{Z_i}^2).$$

The value of the $\delta_{Z_i}$ are the same as those obtained in
the mass renormalization.

After taking into account the different tadpole diagrams which
correct the coupling of the current to one pion states, we find
\begin{eqnarray}
\langle 0|A_\mu^\pm|\pi^\mp(p)\rangle &=&
ip_{\pm\mu}\left[f_\pi-f2\alpha_\pi(I+I_0)\right]\nonumber\\
&&\pm iu_\mu f\alpha_\pi 8J\nonumber\\
\langle 0|A_\mu^0|\pi^0(p)\rangle &=&
ip_{\mu}\left[f_\pi-f4\alpha_\pi I\right]
\end{eqnarray}
with
\begin{eqnarray}
f_\pi &=& f\left(1+\alpha_\pi\bar l_4
\right)
\end{eqnarray}
Now, we can set $f\simeq f_\pi$, $m\simeq m_\pi$ in all ${\cal
O}(\alpha_\pi)$ terms, since any correction will be of order
$\alpha_\pi^2$ (including $\alpha_\pi$), then we define the effective
decay constant as the part proportional to $p_\mu$, so
\begin{eqnarray}
f_{\pi^\pm}(T,\mu_I)&\equiv& f_\pi [1-2\alpha_\pi (I+I_0)]\nonumber\\
f_{\pi^0}(T,\mu_I)&\equiv& f_\pi [1-4\alpha_\pi I]
\end{eqnarray}

\begin{figure}
\includegraphics{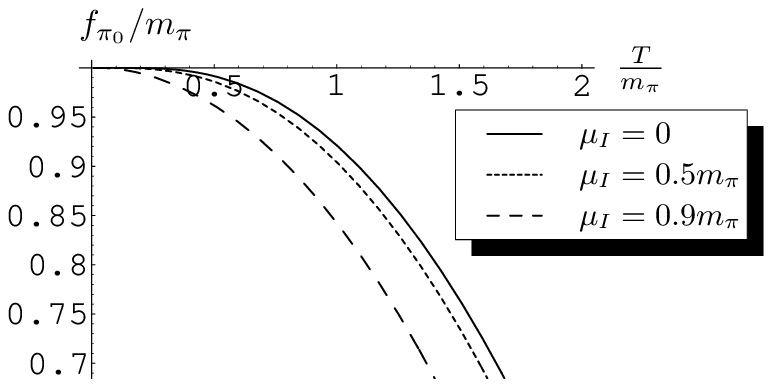}
\caption{ $f_{\pi_0}$ as function of  $T$ for a fixed $\mu_I$ }
\end{figure}

\begin{figure}
\includegraphics{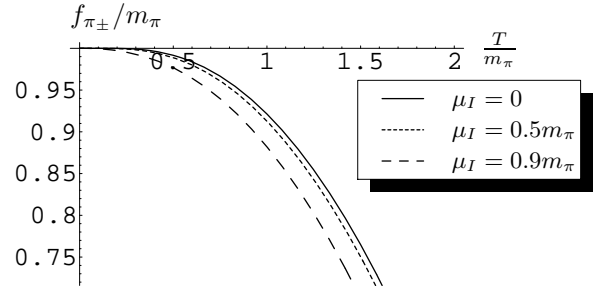}
 \caption{ $f_{\pi^\pm}$ as function of  $T$ for a fixed $\mu_I$ }
 \label{fig4}
\end{figure}

For $\mu _{I}=0$ we agree with the well known results of Gasser and
Leutwyler \cite{GL}. For an increasing finite chemical potential,
 the $f_{\pi }(T)$ couplings decrease faster.
 This effect is enhanced for $f_{\pi _{0}}(T)$ and is related to
  the fact that $f_{\pi _{0}}(T)$only  receives radiative corrections from charged pion tadpoles.

In heavy ion collisions, a finite value of $\mu _{I}$ means that,
 at least locally, we would expect more pions with definite charge
 than in the symmetric case. According to this picture,
  the production rate of dileptons from pion annhilation should be supressed.
   Probably, the detection of such kind of effects will demand a higher center of mass energy.

\begin{figure}
\includegraphics{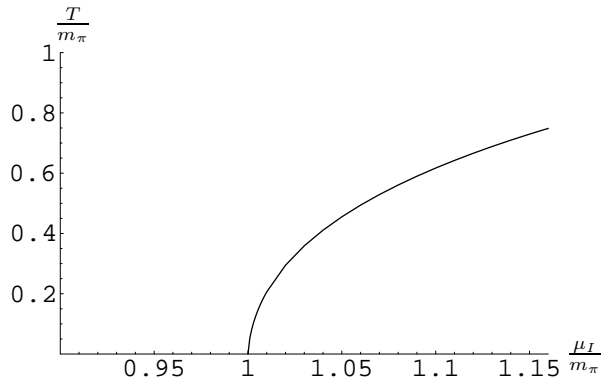}
 \caption{ $T$,$\mu_I$ phase diagram for pion condensation }
 \label{fig5}
\end{figure}

In order to explore the region where $|\mu _{I}| > m_{\pi}$,
associated to a new phase where the condensates occur,
  we need to redefine our fields as fluctuations around the configuration corresponding to a minima
  of the effective potential in ${\cal L}_{2}$. At present we are working on it, but it is possible
  to extrapolate, for $T\ll m_\pi$ and $\mu_I\sim m_\pi$ the condensation point in such a way that
  we actually remain in the first phase. However the curve in the $\mu_I -T$ plane that separates both phases
  is only reliable in the parameters region mentioned before where
  in the thermal factors in eq.(\ref{ints}), we have taken the approximation $n_B(m_\pi x\pm \mu_I) \simeq exp[-\beta(m_\pi x\pm\mu_I)]$.
A complete analysis of the phase can be found in
\cite{splittorff2}. The phase diagram is shown in Fig. \ref{fig5}
in accordance with \cite{son}. However, for higher values of
$\mu_I$ changes abruptly and
  our approximation is no longer valid.\\

\noindent
 {\bf Acknowledgements:}   The work of  M.L. has been
supported
 by Fondecyt (Chile)
under grant No.1010976. C.V. acknowledges support from a Conicyt
Ph.D fellowship (Beca Apoyo Tesis Doctoral).

\end{document}